\documentclass[%
 aip,
 amsmath,amssymb,
 reprint,%
]{revtex4-1}

\usepackage{graphicx}% Include figure files
\usepackage{dcolumn}% Align table columns on decimal point
\usepackage{bm}% bold math
\usepackage[utf8]{inputenc}
\usepackage[T1]{fontenc}
\usepackage{mathptmx}
\usepackage{etoolbox}

\usepackage{xcolor}
\usepackage{hyperref}

\makeatletter
\def\@email#1#2{%
 \endgroup
 \patchcmd{\titleblock@produce}
  {\frontmatter@RRAPformat}
  {\frontmatter@RRAPformat{\produce@RRAP{*#1\href{mailto:#2}{#2}}}\frontmatter@RRAPformat}
  {}{}
}%
\makeatother
\begin{document}

\preprint{AIP/123-QED}

\title[]{A new device for live cell imaging of biological samples under in-plane shear}
% Force line breaks with \\
\author{S.K. Das}
\affiliation{Max Planck Institute for Dynamics and Self-Organization (MPIDS), Am Faßberg 17, 37077 Göttingen, Germany}
 \affiliation{Physics Department, Indian Institute of Science Education and Research (IISER) Pune, Pune 411008, India}
 
 \author{L. Mamoyan}
\affiliation{Max Planck Institute for Dynamics and Self-Organization (MPIDS), Am Faßberg 17, 37077 Göttingen, Germany}

 \author{E. Bodenschatz}
 \affiliation{Max Planck Institute for Dynamics and Self-Organization (MPIDS), Am Faßberg 17, 37077 Göttingen, Germany}
 \affiliation{Faculty of Physics, University of Göttingen, Friedrich-Hund-Platz 1, 37077 Göttingen, Germany}
 \affiliation{Laboratory of Atomic and Solid State Physics, Cornell University, 523 Clark Hall, Ithaca, NY 14853, USA}
 %\altaffiliation[Also at ]{Physics Department, XYZ University.}%Lines breakautomatically or can be forced with \\
\author{S. Villa}%
 \email{stefano.villa@ds.mpg.de.}
\affiliation{Max Planck Institute for Dynamics and Self-Organization (MPIDS), Am Faßberg 17, 37077 Göttingen, Germany}

\date{\today}% It is always \today, today,

\begin{abstract}

\textit{The following article has been accepted by Biophysics Reviews. After it is published, it will be found at \href{https://pubs.aip.org/aip/bpr?_gl=1*jwyd6r*_gcl_au*MTQzMTA3OTg0OC4xNzg1NDAyMDU2}{Link}.}
\\
\\
The effect of mechanical stresses and deformations upon the behavior and properties of cells and tissues is a topic of paramount interest in the field of biophysics. Its understanding is essential for deepening our comprehension of cellular responses to environmental stimuli, with implications in fields ranging from embryogenesis to tumor development. 
For boosting the understanding in this field, it is important to develop devices for the application of simple deformations in order to probe the cellular response to the basic mechanical transformations. While a number of stretching devices have been developed, in-plane shearing deformation has been little investigated. Here, a new device is introduced that allows the application of homogeneous in-plane shear upon a cell monolayer while imaging the sample with a commercial optical microscope. A description of the faced experimental challenges is reported and a full characterization is presented. It is demonstrated the capability of the device to apply reproducible, controlled and homogeneous shear of both single cells and confluent cellular monolayers.

\end{abstract}

\maketitle

\section{Introduction}

From a physical perspective, cells are astonishingly complex out-of-equilibrium systems that not only coordinate their internal processes in order to carry out their functions but are also able to actively respond to external stimuli. They can indeed probe the surrounding environment and actively adapt their properties to it, reacting to both biochemical \cite{helmreich2001biochemistry} and physical\cite{janmey2007cell,hamill2001molecular} stimuli. 

Within a tissue, each cell is mechanically coupled to its neighbors through adherens junctions, desmosomes and the actin myosin cortex \cite{friedl2009collective}. As a result of cell dynamics, both shear and compressive/expansive stresses are therefore generated inside the tissue. In turn, the stresses internal to the monolayer are sensed by the cells and trigger different responses to protect cellular integrity, such as nuclear \cite{villa2022non,frittoli2023tissue} and junctional stiffening \cite{yonemura2010alpha}. Additionally, cell stretch affects cell contractility \cite{kaunas2005cooperative,matthews2006cellular,prado2025molecular} and proliferation \cite{birukov1995stretch,yano2004mechanical,wang2005stretch}. Finally, collective cell migration is in turn affected by these changes in the mechanical properties of cells, in an intricate feedback system \cite{lange2013cell,jain2021mechanical}.
The internal stresses induced by cells motility are not the only ones present in a cellular system. In real tissues, indeed, external deformations due to stretch, compression or shear also contribute.

To improve our understanding of such an intricate interplay, unbind the different contributions and unveil the role of external stresses upon cell phenotypes, it is useful to employ simplified systems. One option is to investigate \textit{in vitro} model systems where an external given deformation – independent from cellular dynamics and mechanical properties – is applied, and to look at how cellular dynamics is affected.
%This is also useful for revealing how constant or repeated mechanical deformations impact on living tissues.
This is particularly relevant given the role of stresses and migration in key biological phenomena such as embryogenesis \cite{agarwal2021mechanosensing,agarwal2021mechanosensing,valet2022mechanical} and tumor progression \cite{friedl2009collective,wells2013targeting,stuelten2018cell,palamidessi2019unjamming,fenner2014macroscopic,lin2015mechanical,wei2016forcing,sauer2023changes}. A deeper understanding of the role of external macroscopic deformations in triggering a change in cell properties (migration, stiffness, gene expression, ...) could lead to new insights into these pivotal biological processes.

In recent years, a number of experimental devices have been developed to enable a direct imaging of microscopic rearrangements within soft matter samples while an external stress or deformation is applied \cite{aime2016stress,villa2022quantitative,kalafatakis2026shearview,brooks1999interfacial,keim2013yielding}. These systems allow the study of the correlation of the mechanical properties of materials with the inner structure and dynamics of their microscopic components. In the field of biophysics, analogous setups have been developed to apply in-plane stretches to cell monolayers while observing their effects on cell morphology and dynamics. These systems probe either floating monolayers \cite{harris2013generating} or monolayers growing on elastic substrates, such as polydimethylsiloxane (PDMS) \cite{bodenschatz2022epithelial}. Published works investigating such systems highlighted how shear directly affects the average cell projected area, with a significative impact on both cell contractility and cell proliferation\cite{monemian2019tissue,riehl2012mechanical,clause2009engineered,xie2020mechanical}. More in detail, when subjected to extensile deformation, cells react to dissipate the stresses in different ways. A first strategy is to align themselves along the direction of the load, as it has been documented for a $39\%$ strain applied to MDCKII cells\cite{harris2013generating}. A second relaxation mechanism is an oriented cell duplication along the strain direction, as it has been documented for MDCK cells upon the application of an uniaxial strain\cite{wyatt2015emergence}. This phenomenon has been related to an active response of the cells trough a reorganization of the cytoskeleton for a $12\%$ uniaxial strain \cite{hart2017cadherin}.
Stretch also has been found to impact on the dynamics of the monolayer: a cyclic $10 \%$ uniaxial stretch reduces the average motility of epithelial cells, with a stronger impact for lower stretching frequencies (reduction of an approximate factor $1.5$ for a $0.1$ Hz stretch)\cite{dow2024cyclic}. Recruitment of vinculin to stabilize the adherens junction has been found to be responsible for this reduction in the dynamics.
Additionally, cell monolayers exhibit non-trivial changes in their mechanical properties under stretch and stress. Stretch of less than $10\%$ of lung epithelial cells results in a remarkable increase of the viscoelastic response of the cells due to the remodeling of the actin cytoskeleton\cite{trepat2004viscoelasticity}. At the same time, pressure induced bi-axial stretching of monolayers of kidney epithelial showed evidence of strain softening. This mechanism, triggered by a stretch-induced dilution of the actin cortex, enables epithelial monolayers to undergo large stretching under constant tension, with cells reaching a more than 5-fold increase in their projected area \cite{latorre2018active}.

A different and less explored type of deformation is the in-plane shear deformation. Contrary to the apico-basal shear flow deformation, the in-plane shear imposes a strain to the cells in the same plane of the monolayer. Differently from stretch, however, it is a deformation that preserves the average cellular projected area. Compared to stretching, a shearing deformation moreover reduces the amount of stress applied to adherent junctions normal to the cell membrane and increases the ones tangential to it. This is expected to differently impact on the mechanical evolution of the junctions\cite{kale2018distinct} and thus on the mechanical and dynamic properties of the monolayer.
Despite the relevance of this type of deformation, up to now in-plane shear has only been explored by the work by Sadeghipour et al. in 2018 \cite{sadeghipour2018shear}. There, a cell monolayer was cultured on two adjacent glass slides. One slide was then displaced parallel to the common edge, shearing the cells that were attached along the border between the slides. A remarkable increase in the dynamics has been observed following the shearing, with the emergence of a collective wave toward the shear-plane. This pioneering work proved the relevance of shear deformations for epithelium motility. However, the shear was limited to a single line of cells and not to the entire 2D plane of the experiment. Currently only simulation works have explored the latter condition, predicting dramatic effects on cellular dynamics \cite{huang2022shear}.

Here, a new device is introduced and characterized that enables the application of uniform, in-plane shear deformation to 2D samples. This device is suitable for applying deformations to both single cells and confluent cell layers, allowing a simultaneous imaging of the sample, as it is designed to be compatible with commercial optical microscopes. This tool will therefore enable a systematic study of the effect of such deformation, hitherto unexplored, on cells and tissues.

\section{Design of the device}

The structure of the shearing device is mostly made of aluminum and is designed to be mounted on commercial microscopes. It mounts an elastic chamber made of PDMS, specifically designed for this application. 

\subsection{PDMS chamber}
\begin{figure}
\includegraphics[width=0.5\textwidth]{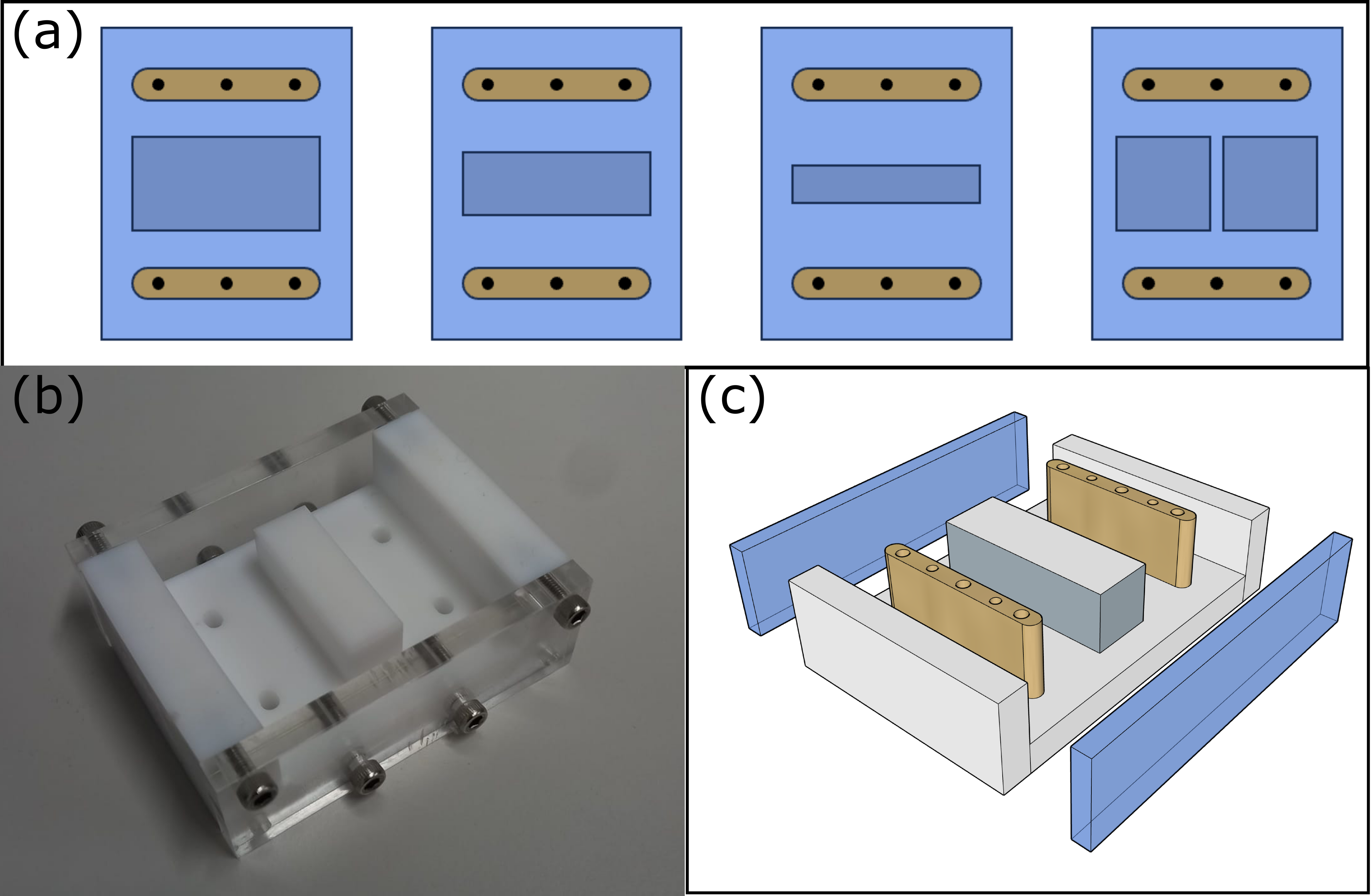}% Here is how to import EPS art
\caption{\label{fig:PDMSchamber} (a) Examples of different designs for the PDMS chambers where the frame (light blue), the membrane (blue), and the two plastic holders (ocher) are visible. From left to right: $3$ cm x $1.5$ cm chamber (pre-stretch $13\%$), $3$ cm x $1$ cm chamber (pre-stretch $20\%$), $3$ cm x $0.6$ cm chamber (pre-stretch $33\%$), and two-wells $1.5$ cm x $1.5$ cm chamber (pre-stretch $13\%$). (b) Mold for the production of the $3$ cm x $1$ cm chamber. (c) Sketch of the mold reported in (b), where the position of the plastic holders (ocher) is visible.}
\end{figure}

In order to apply the shear to the sample, cells must be seeded on an elastic substrate. For this, a chamber made of Polydimethylsiloxane (PDMS) with one or more wells has been developed. The viscous response of PDMS is negligible at the characteristic time scales explored in the shearing protocol considered here \cite{darby2022modulus,villa2022quantitative} and can thus be considered a purely elastic material that propagates the mechanically applied strain to the cell monolayer without any delay. Additionally, PDMS enables the optical access, thus allowing imaging of the samples both in transmission (\textit{e.g.}, for bright field, phase contrast or DIC microscopy) and in reflection (\textit{e.g.}, for epifluorescence or confocal microscopy). The PDMS chambers (Fig. \ref{fig:PDMSchamber}(a) ) consist of a rectangular frame $1$ cm thick embedding one or more central wells and two plastic holders produced with a 3D printer (HP Jet Fusion 540 3D Printer). The wells are closed on the bottom by a thin PDMS membrane to which cells can adhere while immersed in their culture medium. The plastic holders are located at the two ends of the long side of the rectangular frame and are used to mount the chamber on the shearing device. For this, they have 3 passing holes that fit two sets of three pins present on the shearing device. Tests revealed that 3D printed plastic holders are much more efficient than plastic holders obtained from other manufacturing processes like milling. This is because the intrinsic roughness of the 3D printed material improves the grip of PDMS on the plastic holder, greatly reducing the events of loss of adhesion between the two materials.

The chamber is produced in two steps, as illustrated in Fig. \ref{fig:PDMSprepSteps} of the Appendix. First, the external frame is produced by mixing the silicon oil (Sylgard\textsuperscript{TM} 184 Silicone Elastomer Base) and the crosslinker (Sylgard\textsuperscript{TM} 184 Silicone Elastomer Curing Agent) at a mixing ratio $25$:$1$ for a total mass of $15$ g. The mixture is then stirred with a stick and placed in a vacuum desiccator ($\sim 0.002$ mbar) for $10$ minutes in order to remove the air bubbles. The de-bubbled sample is then poured into a Teflon mold (Fig. \ref{fig:PDMSchamber}(b)-(c)). The two plastic holders are screwed to the mold from the bottom side before pouring the PDMS mixture in order to keep them in place while pouring and curing the PDMS. The mold is then placed in an oven pre-heated at $70^\circ$C for $90$ minutes. After the treatment, the PDMS frame embedding the plastic holders is removed from the mold. The low PDMS-Teflon adhesion facilitates the detachment.
The membrane is prepared by mixing the PDMS base and curing agent at a mixing ratio $10$:$1$, stirring and debubbling as for the preparation of the frame. The mixture is then poured on a silicon wafer and spin coated for $20$ s at $800$ RPM, in order to produce a layer of PDMS with regular thickness of $100$ \textmu{}m. Such extension is sufficient for preventing fractures upon stretch and shear while avoiding excessive thickness that could impair the optical access to the sample.
After the spin coating step, the silicon wafer is placed in the oven pre-heated at $70^\circ$C for $45$ minutes. A second layer of PDMS is then poured and spin coated as the first one. Before curing the second layer, the frame is gently placed on the uncured PDMS. Gentleness is needed to prevent the sliding of the frame on the silicon wafer once in contact with the PDMS, thus avoiding thickness inhomogeneities in the latter. Finally, the silicon wafer with the PDMS frame is placed in the oven pre-heated at $70^\circ$C for $45$ minutes for curing the second layer of the membrane and join the frame and membrane. A cutter is finally used to cut out the shape of the membrane along the edges of the frame and detach the chamber from the silicon wafer. The $10$:$1$ mixing ratio maximizes the stiffness of the cured PDMS\cite{darby2022modulus}. This also reduces the stickiness of the cured PDMS, improving the detachment from the silicon wafer. %, and contributes to prevent the formation of wrinkles during the shear (TRUE?...metti ref).
The low thickness enables the imaging with short working distance objectives for inverted microscopy setups. For even shorter working distances, the spin coating protocol can be easily adapted to obtain thinner PDMS membranes.

As it can be seen in Figure \ref{fig:PDMSchamber}(a), different well shapes are possible by preparing different molds. Here it has been chosen to only consider rectangular (or square) shapes to have uniform shear across the sample. For probing different cellular conditions at the same time, a multiwell option is also possible (Fig. \ref{fig:PDMSchamber}(a)).

\subsubsection{Wrinkles}
\begin{figure}
\includegraphics[width=0.5\textwidth]{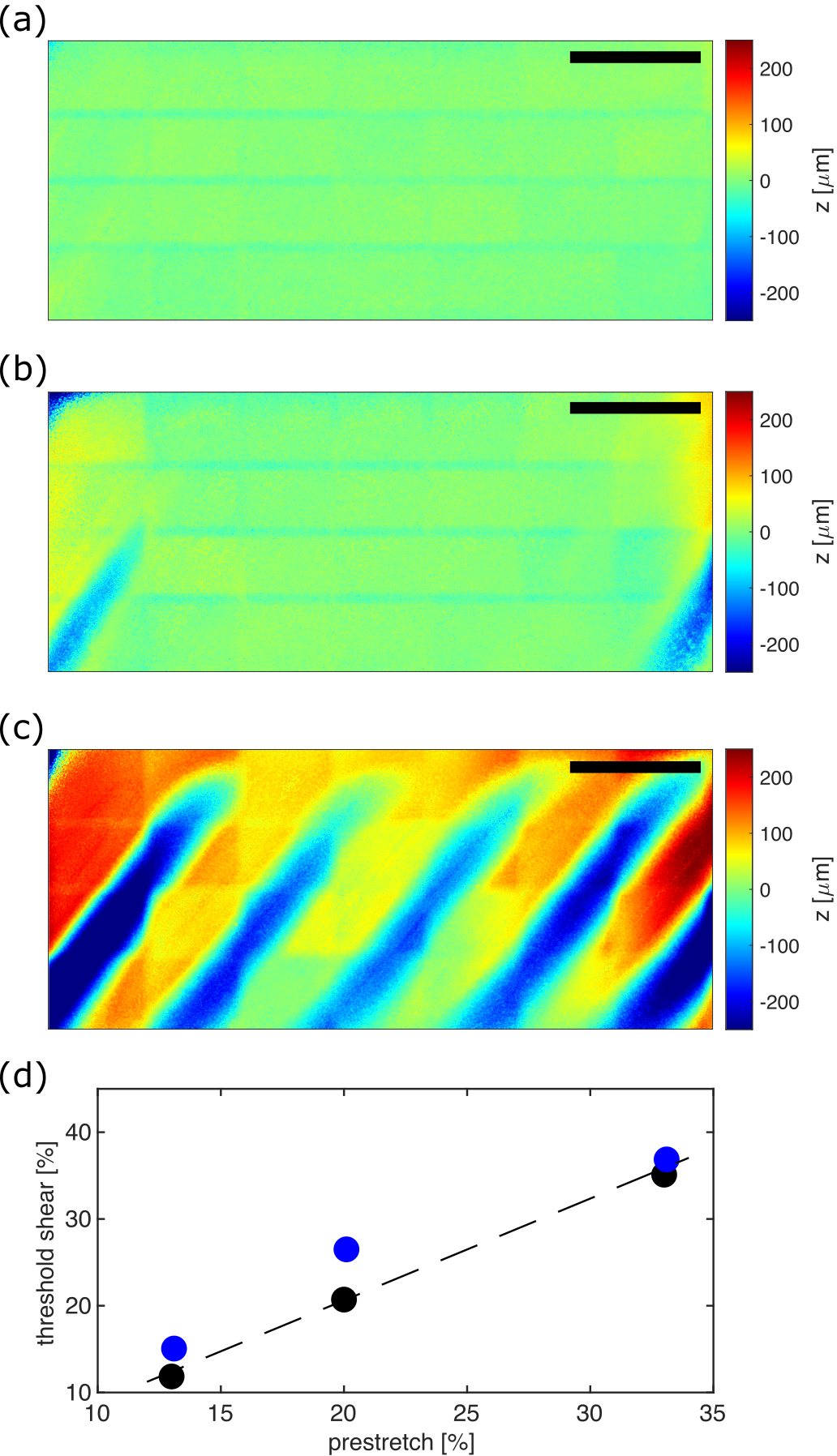}% Here is how to import EPS art
\caption{\label{fig:wrinkles} (a-c) Profilometer data representing the whole PDMS chamber for a $20\%$ pre-stretch, with an effective applied strain of $23.4\%$, $29.6\%$ and $36.6\%$, respectively. It is possible to see the no wrinkle case (a), the presence of a flat region in the center of the membrane between the lateral wrinkles (b) and pervasive wrinkles case (c). The scale bar is $5$ mm. (d) Shear thresholds at which lateral wrinkles (black) and pervasive wrinkles (blue) appear for different applied pre-stretches. The dashed line is a linear fit of the black points. Please note that the reported shear is the measured effective one and not the nominal applied shear (see section \ref{sec:effectiveshear}).}
\end{figure}

The produced chamber can be in principle sheared at strains larger than $100\%$, up to its breaking point. However, wrinkles can form on the PDMS membrane upon shear, thus affecting the sample flatness (Fig. \ref{fig:wrinkles}(a)-(c)). This introduces a non-ideality in the deformation field of the PDMS and thus of the cells. Moreover, presence of curvatures may induce anisotropy in cellular dynamics due to preferential curvatures along which cells may spread and move when adherent on a substrate that is not planar (curvotaxis) \cite{pieuchot2018curvotaxis,sadhu2024minimal}. As a consequence, reliable  results can only be obtained when wrinkles are not formed.
The formation of wrinkles is due to the membrane buckling originating from its inability to withstand compressive loading \cite{wong2006wrinkled,iwasa2018wrinkle}. Consequently, the application of a pre-stretch in the direction orthogonal to the shear is expected to reduce the appearance of wrinkles \cite{luo2017wrinkle}. In particular, a linear relationship is predicted between the applied pre-stretch and the threshold shear at which the wrinkles form \cite{iwasa2018wrinkle}.

The PDMS chambers are therefore produced $2$ mm shorter than the shearing device in order to apply a pre-stretch on the membrane. When the chamber is mounted on the shearing device, indeed, the gap in length translates in an effective pre-stretch whose amount depends on the length $Y$ of the rectangular well in the pre-stretching direction ($y$).  In the present work, three different values of $Y$ have been chosen ($0.6$ mm, $1.0$ mm and $1.5$ mm), corresponding to pre-stretches of $13\%$, $20\%$ and $33\%$.

The 3D shape of wrinkles has been measured using a laser profilometer (Keyence VK-X200). To enhance the reflectivity of the membrane, for this measurement the membrane has been covered by a monolayer of dry glass beads (diameter $10$ \textmu{}m). In Figure \ref{fig:wrinkles}(a)-(c) three examples are reported for the same chamber sheared at different amplitudes. For the lowest shear amplitude, no wrinkles are visible (Fig. \ref{fig:wrinkles}(a)). At intermediate shear, wrinkles emerge at the lateral sides of the membrane (Fig. \ref{fig:wrinkles}(b)), with a residual flat region in the center of the chamber. Here and in the following, wrinkles are considered lateral and not pervasive when a central region of the membrane large at least as its height is not occupied by wrinkles. For higher shearing amplitudes, then, periodic wrinkles pervade the entire membrane (Fig. \ref{fig:wrinkles}(c)). This behavior with shear is in qualitative agreement with numerical predictions \cite{iwasa2018wrinkle}. 

In Figure \ref{fig:wrinkles}(d) it is reported the output of a systematic test of the threshold shears at which the wrinkles appear at the sides of the membrane (black points) and at which they pervade the entire chamber (blue points) for different applied pre-stretches. For shear lower than the black points, the membrane is flat, while for intermediate shear between blue and black points, only the central region of the membrane can be safely considered flat. For larger deformation, finally, the whole sample is affected by wrinkles. 
The wrinkle emergence thresholds appear to be fairly linear with the applied pre-stretch, in agreement with numerical predictions \cite{iwasa2018wrinkle}.
 Using the $33\%$ pre-stretch chamber it is therefore possible to reach 
 an effective shear strain up to $40\%$ without affecting the sample planarity. In order to reach larger deformations, a chamber with higher pre-stretch is required.

\subsubsection{Pre-stretching system}

\begin{figure}
\includegraphics[width=0.5\textwidth]{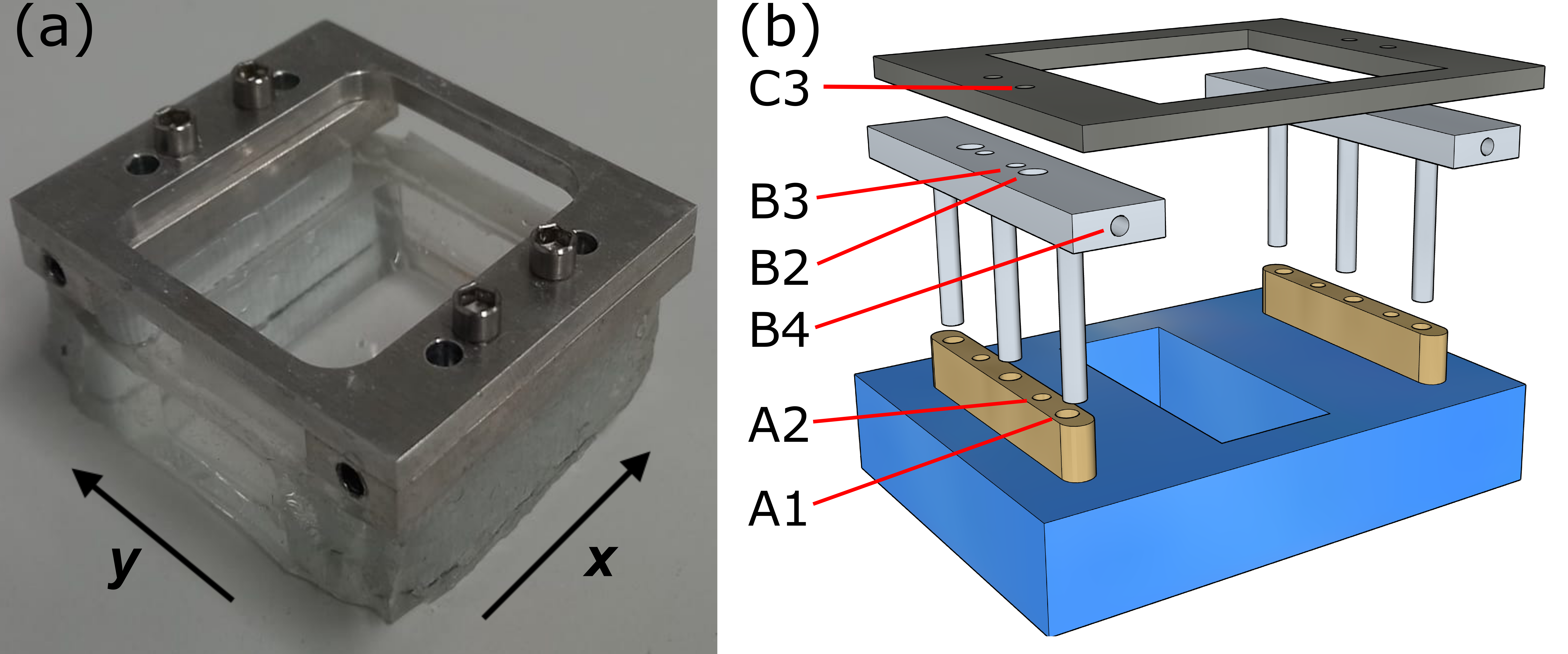}% Here is how to import EPS art
\caption{\label{fig:PDMSprestretcher} (a) Picture of a PDMS chamber mounted on the pre-stretcher. (b) Sketch illustrating the different parts of the pre-stretcher: the pre-stretcher frame (C), and the aluminum parallelepipeds (B). The latter mount the pins that fit the holes in the plastic holders (A). Three sets of holes are present on B for connecting it to the plastic holders (B2 to A2), to the pre-stretcher frame (C3 to B3), and to the shearing device (B4).}
\end{figure}

The stretch of the PDMS chamber that is needed to prevent wrinkle formation under shear causes an increase of the surface area of the PDMS membrane.
 For samples requiring long preparations under environmental control (\textit{e.g.} cell culture in an incubator) before the experiment, it is therefore important to keep the same stretching of the chamber before mounting it on the shearing device, in order to avoid imposing a stretch to the cells when subsequently positioning the sample on the setup. Deformation of the PDMS membrane during the process can indeed affect, or even damage or detach, the cells. To this end, after preparation the PDMS chambers are mounted on a pre-stretcher consisting of three parts (Fig. \ref{fig:PDMSprestretcher}). Referring to Figure \ref{fig:PDMSprestretcher}(b): two aluminum parallelepipeds (B) mount three pins each that fit the holes present in the plastic holders (A). Two passing through holes (B2 to A2) allow to firmly attach the plastic holders (and thus the chamber) to the aluminum parallelepipeds. The two parallelepipeds are then screwed to an aluminum frame (C) that imposes to the chamber the same pre-stretch of the shearing device.

\subsection{Shearing device}
\begin{figure*}
\includegraphics[width=1\textwidth]{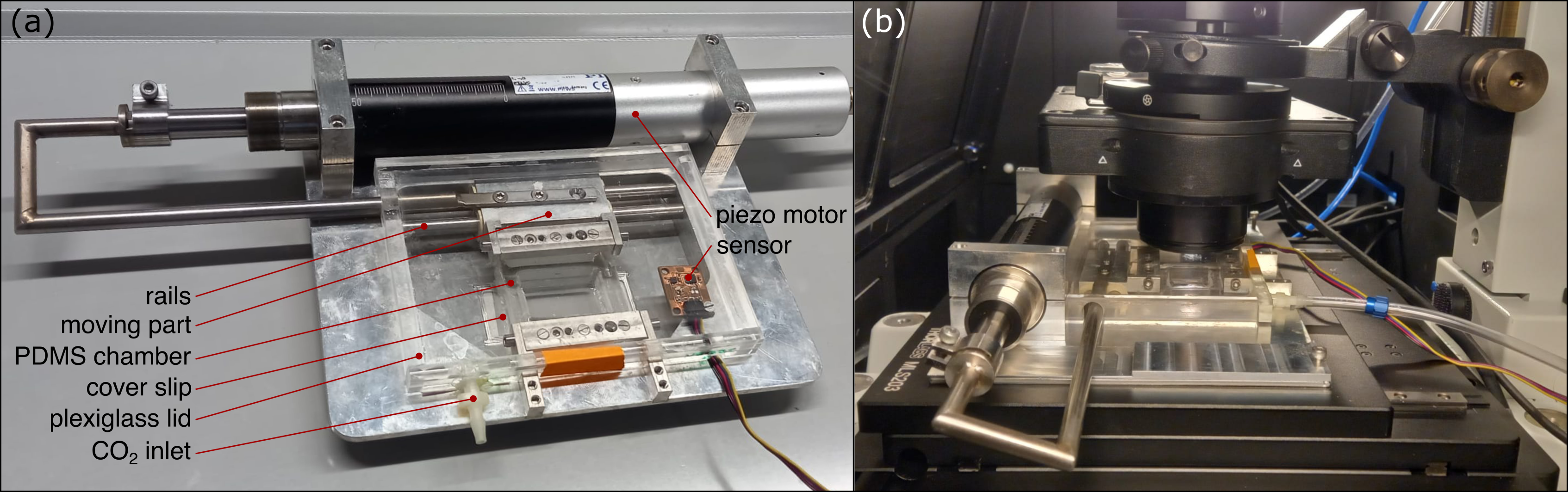}% Here is how to import EPS art
\caption{\label{fig:shdev} (a) Picture of the shearing device with the names of its main parts. (b) Picture of the shearing device mounted on the sample stage of an Olympus IX81 inverted microscope mounting a Pecon incubator. }
\end{figure*}

The shearing device has been designed to be compatible with commercial microscopes mounting an environmental chamber for live cell imaging. It has been tested on Olympus inverted microscopes IX81 and IX83, both mounting a Pecon incubator (Fig. \ref{fig:shdev}(b)) , and on a Nikon Eclipse Ti2 inverted microscope mounting an Okolab H203-ENCLOSURE climate chamber (Fig. \ref{fig:nikon}). It can be easily adapted to any inverted microscope mounting a large chamber incubator.

The shearing device is shown in Figure \ref{fig:shdev}. It consists of an aluminum frame mounting two cylindrical rails, on which a moving block is free to slide. The two opposite sides of the PDMS chamber are mounted on the moving block and on the aluminum frame, respectively. Once the PDMS chamber is mounted, by translating the moving block a shear is applied to the PDMS chamber and thus to the membrane and the cells adherent to it\footnote{Please note that because of this mechanism for propagating the deformation and because of the PDMS stiffness much higher than the one of cells, the device here introduced can only be employed for a strain-controlled shear, and not for a stress-controlled one.}. The moving block is controlled by a piezoelectric motor (PI M-235.5DD High-Resolution Linear Actuator) allowing a maximum displacement range of $5$ cm, with a minimum step size of $1.5$ \textmu{}m and a maximum speed of $30$ mm/s. The motor can be controlled both via the PI software and via programming languages such as Matlab. A Matlab code has been developed to control the moving block with different protocols (step shear, square wave, sinusoidal wave,...), thus applying to the sample the desired shearing protocol.

Most of live \textit{in vitro} experiments with biological samples require a specific temperature and $CO_2$ supply to guarantee cell survival, typically $37^\circ$C and a $5\%$ concentration of $CO_2$. Additionally, humidity around the sample must be controlled for long time-scale experiments to prevent the evaporation of the cellular medium. Temperature control is provided by the large commercial climate chamber. In order to control $CO_2$ concentration and humidity, the PDMS chamber, the moving block and the rails of the shearing device are confined in a closed space, delimited on the bottom by an aluminum plate with a window closed by a coverslip in correspondence of the PDMS chamber. On the top, the same region is enclosed by a plexiglas lid with a side inlet for $CO_2$ and humidity control. The latter are provided externally, from the environmental control of the microscope climate chamber unit.

For systems where a large chamber incubator system is not available, it is possible to replace the glass coverslip closing the imaging window with an indium tin oxide coated glass slide and implement a heating system by externally imposing a voltage to it with a power supply.

Due to the importance for cell viability of providing a precise environmental control, a sensor has been developed to measure the temperature, the humidity and the $CO_2$ concentration. 
A sensor for $CO_2$ concentration (Sensirion STC31) and a sensor for temperature and humidity (Sensirion SHTC3) have been assembled on a printed circuit board (PCB) mounted on the shearing device below the plexiglass lid. The PCB is wired to a microcontroller board (Arduino Uno Rev3), coded in C++ to download the data to a computer and display the environmental parameters on a LCD screen (Joy-it SBC-LCD16x2) for easy viewing during experiments.

\section{Sample preparation}

\subsubsection{Chemical treatment for improving cell adhesion}

After fabrication, the PDMS chamber is carefully cleaned by removing solid fragments with compressed air and washing the well with Dulbecco's Phosphate Buffered Saline (PBS, Thermo Fischer Scientific catalog number 14190144). The chamber is then sterilized using the UV light of a biological hood for $30$ minutes. 
The PDMS chamber can then be directly used for experiments with samples with good adhesion on a wide range of materials, like \textit{Dictyostelium discoideum} (Dd), which do not form focal adhesion \cite{tarantola2014cell}. For many cell lines, including mammalian epithelial cells, however, the PDMS hydrophobicity makes cell adhesion difficult and a chemical treatment of the membrane must be made before seeding the cells on the substrate.
For this, a protocol adapted from Siddique \textit{et al.} 2017\cite{siddique2017improved} has been chosen and tested. An $O_2$ plasma etching is first performed for $2$ minutes at $50\%$ power (Plasma-System Pico, generator power $200$ W). Immediately after that, a $2\%$ (3-Aminopropyl)triethoxysilane (APTES, Sigma-Aldrich 440140) solution in acetone is poured into the well. After $15$ minutes, the APTES solution is removed and the well is washed $3$ times with PBS. Visual inspection immediately reveals a decrease of the contact angle of PBS on the PDMS membrane. Drop measurement shown a change in the contact angle of water on PDMS from $90^\circ$ to $46^\circ$ following the APTES treatment, thus attesting to the increased hydrophilicity of the PDMS membrane.

In order to check the effect of PDMS substrate upon cell dynamics, an untreated $35$ mm plastic petri dish has also been coated with a thin layer of PDMS (10:1) obtained by simply pouring uncured PDMS in the petri dish and curing it in the pre-heated oven as for the PDMS frame. The PDMS layer has then been treated with the same APTES procedure described for the PDMS membrane.

\subsubsection{Substrate coating and cell preparation}

\textbf{Fibronectin coating.}
Depending on the sample and on the experiment, a coating for more fine tuning cell-substrate adhesion can be performed. For the MDCKII cells used in the present paper, a Fibronectin (Sigma F1141 bovine plasma fibronectin) coating has then been made as it follows. 
The PDMS well is first washed $3$ times with PBS. Then a fibronectin solution in distilled water ($0.1$ mg/ml) is added in the well. After $1$ hour incubation at room temperature, the fibronectin solution is removed, the well is washed three times with PBS and is left to dry under a biological hood for $45$ minutes. If the chamber is not used immediately, it can then be stored for few days in a $4$ C$^\circ$ fridge after adding PBS in the well. 
The same coating protocol has been applied to petri dishes with and without PDMS coating used for tests at rest.

\textbf{MDCKII culture.}
Madin-Darby Canine Kidney cells (MDCKII) were kindly provided by the group of Prof. Andreas Janshoff (University of Göttingen, Germany). They are cultured in MEM (Minimal essential medium with Earl’s salts, $2.2$ g/L NaHCO\textsubscript{3}, and 2 \textmu{}M GlutaMAX, ThermoFisher Scientific, Waltham, Massachusetts, USA) with $10\%$ FCS (Fetal Bovine Serum, BioWest, Nuaillé, France) at $37^\circ$C and $5\%$ CO\textsubscript{2} in a humidified incubator. 
After the fibronectin coating, $3 \cdot 10^4$ cells/cm\textsuperscript{2} are seeded on the PDMS membrane or in the prepared petri dishes. Confluency is reached after $48$ hours.

\textbf{\textit{Dictyostelium discoideum} culture.} 
\textit{Dictyostelyium discoideum} ACA Null AX2 cells (D.d.) were grown in HL-5 medium ($35.5$g of Formedium powder from Formedium Ltd., England, per litre of double-distilled water, autoclaved and filtered) at $22^\circ$ on polystyrene Petri dishes (TC Dish 100, Sarsted, Germany) and harvested when
 they became confluent. Before the experiments, the cells were centrifuged and washed twice with $10$ mL SorC buffer ($8.0$g KH2PO4, $1.16$ g Na\textsubscript{2}HPO\textsubscript{4}, and $4$ mL of $50$ mM CaCl\textsubscript{2} for $4$ L of buffer).
 The cells were then centrifuged at $1000$ RPM for three minutes and resuspended in $1$ mL of Bonner salt ($0.6$ g NaCl, $0.75$ g KCl, $0.3$ g  CaCl2 per litre). The cell density was determined using a hemocytometer (Neubauer Z\"{a}hlkammer), diluted and then plated in the PDMS chamber with a density of $9.25$x$10^4$ cells/cm\textsuperscript{2}.

\section{Characterization of the device}

\subsection{Effective shear}
\label{sec:effectiveshear}

\begin{figure}
\includegraphics[width=0.5\textwidth]{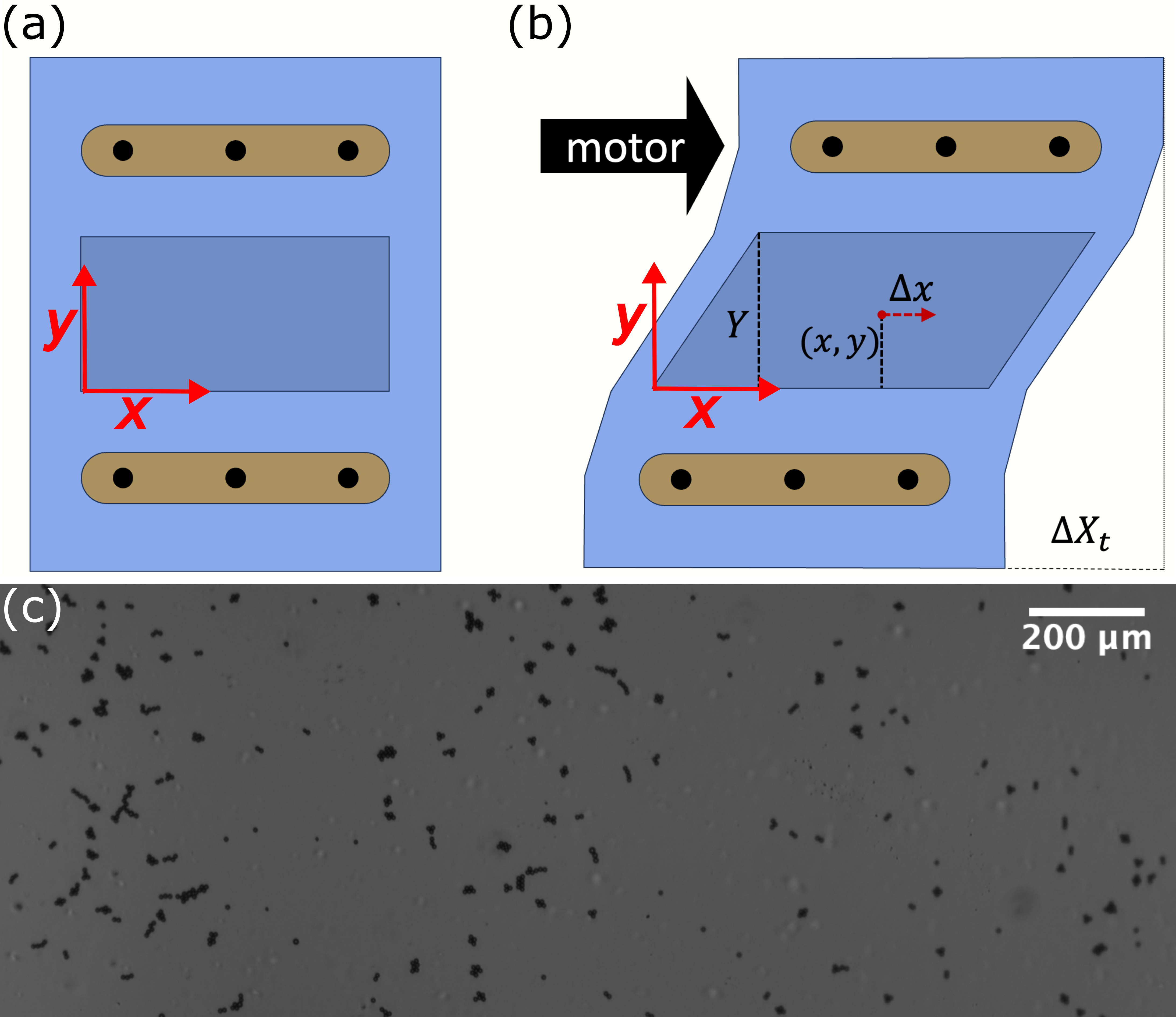}% Here is how to import EPS art
\caption{\label{fig:PDMSshear} Sketch of the PDMS chamber (a) at rest and (b) during the application of a nominal shear $\gamma_n=\Delta X_t/Y$. (c) Microscopy image (4x objective) of the beads embedded in the PDMS membrane for the effective shear characterization.}
\end{figure}

For quantitative measurements on the effect of shear on cell populations, a precise evaluation of the applied deformation is needed. A full characterization has been performed by applying a nominal strain to an empty PDMS chamber and measuring the effective shear. 
For notation clarity, the Cartesian coordinates ($x,y$) are introduced, where $x$ is the shearing direction and $y$ is the one orthogonal to it. The axes origin is chosen at the bottom-left corner of the PDMS membrane, as shown in Figure \ref{fig:PDMSshear}, and rigidly moves with it during shear. The nominal strain is defined as $\gamma_n=\Delta X_t/Y$, where $Y$ is the size of the membrane in the direction $y$ and $\Delta X_t$ is the relative displacement along $x$ of the moving plastic holder with respect to the fixed one. It corresponds therefore to the strain obtained assuming that the central part of the PDMS (the membrane) is the only part that deforms during shear. Indeed, for the same globally applied strain the PDMS membrane deforms more than the frame because of the difference in PDMS thickness%($100$ \textmu{}m versus $1$ cm)
, resulting in a difference in the effective elastic constant. % $k_f$ much higher than the effective stiffness $k_s$ in the central region. 
For a given $\gamma_n$, the local effective strain has been evaluated as  
$\gamma(x,y)=\Delta x(x,y)/y$, where $\Delta x(x,y)$ is the effective displacement during shear along $x$ of a point $(x,y)$ on the membrane.

In order to characterize the response of the chamber to the shear application, a systematic study has been performed of the local effective shear $\gamma$ for a given $\gamma_n$ .
For tracking displacements on the membrane, a PDMS chamber has been prepared with a layer of $8.1$ \textmu{}m diameter colloids embedded between the two PDMS membrane layers (Fig. \ref{fig:PDMSshear}(c)). The chamber was mounted on the shearing device and imaged under a bright-field microscope at 4x magnification. Images have been recorded at $2$ frames per second (fps) while applying a step shear deformation for different nominal shear amplitudes, and the beads displacement has been quantified using space correlation. %($6.9\%$, $10.3\%$, $13.8\%$, and $17\%$).
For each nominal shear, the imaged deformation has been repeated upon different field of views (FOVs) in order to collect the displacements in the entire PDMS membrane region. Each $3$x$3$ mm\textsuperscript{2} field of view has then been divided in four equal regions of interests (ROIs). For each ROI centered at coordinates $(x,y)$ the corresponding displacement $\Delta X$ has been quantified as the average displacement of the beads contained in it.

\begin{figure}
\includegraphics[width=0.5\textwidth]{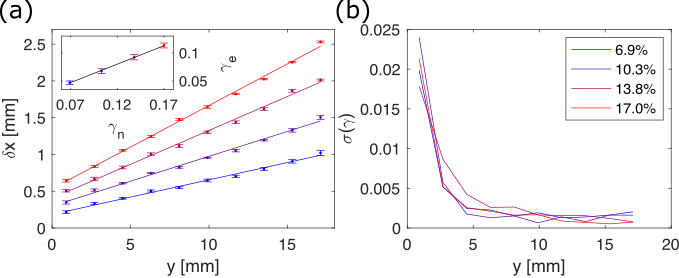}% Here is how to import EPS art
\caption{\label{fig:effShear} (a) Measured PDMS membrane deformation $\delta x$ as a function of $y$ for different nominal shears (from blue to red: $6.9\%$, $10.3\%$, $13.8\%$ and $17.0\%$) and the corresponding linear fits. In the inset are reported the fitted slopes of $\delta x$ (effective shear $\gamma_e$) as a function of the nominal shear together with a linear fit (black line). The latter has a fitted slope equal to $0.65\pm 0.07$. (b) Standard deviation of the local strain, $\gamma$, versus $y$. For the present characterization, a $13\%$ pre-stretch chamber has been used.}
\end{figure}

In Figure \ref{fig:effShear}(a) is reported the average displacement $\delta x = <\Delta x \left(x,y\right)>_x$ as a function of the $y$ coordinate for different nominal shears. Here, $<...>_x$ denotes the average over $x$ within the limits of the membrane. From the linear fit of those points the global effective shear $\gamma_e$ is obtained and reported in the inset. As it can be seen, a linear relationship occurs between $\gamma_n$ and $\gamma_e$, with a slope equal to $0.65 \pm 0.07$ and a negative intersection with the horizontal axis. Consequently, the effective shear appears to be about the $65\%$ of the nominal shear, regardless its value. With this calibration, it is possible to reliably apply the desired effective shear to a sample in the characterized chamber without the need of a direct control of the global deformation. 
The reason for such a large discrepancy between nominal and effective strain is the deformation the PDMS frame also undergoes: the nominal applied strain is applied to the entire frame, and not only to the membrane, and thus the total strain is distributed across the frame and the membrane. This also explains the negative intersection between the fits reported in Fig. \ref{fig:effShear}(a) and the horizontal axis. Different chambers prepared with the described protocol are not expected to differ much in thickness. However, for a precise control of the shear applied to the sample a characterization of the ratio between effective and nominal shear should be performed for each prepared chamber. Given the linear relation between nominal and effective strain, this can be simply done tracking the displacements of the four angles of the PDMS membrane upon the application of a given shear.
In order to test the homogeneity of the deformation, in Figure \ref{fig:effShear}(b) it is reported the local shear strain standard deviation $\sigma_\gamma = <[\Delta x \left(x,y\right) - \delta x\left(y\right)]^2>_x/y$ as a function of $y$ for different nominal shears. First, $\sigma_\gamma$ appears to be independent from the strain. Moreover, at the side of the PDMS membrane closer to the fix grabbing point the deviation is of the order of $2\%$, but as soon as $y$ increases, it drops below $0.3\%$. This is because the standard deviation of the displacement is rather constant with $y$ (see Appendix \ref{app:dx_vs_x}) and thus the standard deviation of $\gamma$ scales with $y^{-1}$. In conclusion, the deformation is strongly homogeneous in most of the PDMS membrane.

\subsection{Tests with cells}

In order to test the capability of the shearing device in applying in-plane shear to cellular samples, a test has been made of shearing individual (\textit{Dictyostelium discoideum}, D.d.) and confluent (MDCKII) cells. The former are ameboid cells that do not form focal adhesion. Conversely, MDCKII is a cell line of epithelial cells, forming monolayers with strong inter-cellular adhesion and the formation of focal points on the substrate. 

\begin{figure*}
\includegraphics[width=0.8\textwidth]{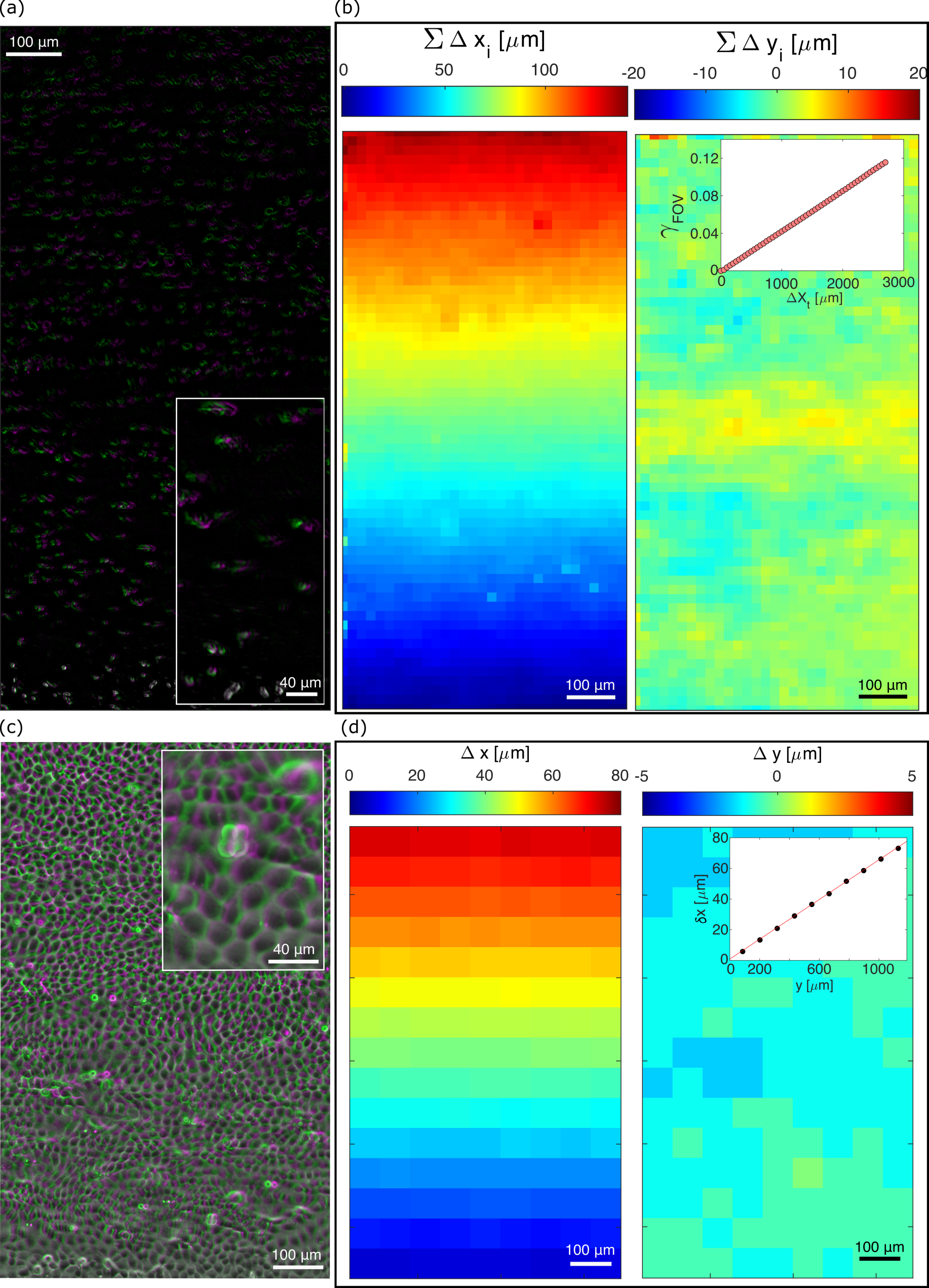}% Here is how to import EPS art
\caption{\label{fig:cells} (a) Overlap of two frames from the D.d. experiment, corresponding to shear amplitudes $\gamma_e=0\%$ (green channel) and $\gamma_e=4.2\%$ (magenta channel). In the inset, a zoom on a detail. (b) Cumulative sum along $x$ (left) and $y$ (right) of the displacements resulting from the application of a $12.8\%$ shear over the same FOVs of (a).  In the inset of (b) it is reported the cumulative shear strain averaged over the entire FOVs reported in (a) as a function of the motor displacement. (c) Overlap of two frames from the MDCKII experiment corresponding to shear amplitudes $\gamma_e=0\%$ (green channel) and $\gamma_e=6.5\%$ (magenta channel). (d) Quantified displacements along $x$ (left) and $y$ (right) resulting from the application of a $6.5\%$ shear over the same FOVs of (c). In the inset of (d) it is reported $\delta x=<\Delta x>_x$ versus $y$ relative to the same FOV and applied strain of (c). The red line is a linear fit of the displacements resulting from the shear application. Both samples are prepared in a $20\%$ pre-stretch chamber and are imaged in a region close to the center of the membrane.}
\end{figure*}

A first test has been performed with a chamber containing D.d. cells. These cells are rather simple to handle, as they survive at room temperature without requiring an external CO\textsubscript{2} supply. They have been seeded in a PDMS chamber, mounted on the shearing device and imaged in phase contrast with a 10x objective. Then an overall displacement of $3.4$ mm has been applied to the moving part of the shearing device (corresponding to $\gamma_e=12.8\%$) at $50$ \textmu{}m steps. After each step, an image has been taken of the chosen field of view (FOV), where D.d. cells were visible. In Figure \ref{fig:cells}(a) two of these frames are overlapped. 
The overlap is made by performing a rigid translation along $x$, in order to overlap the cells in the bottom side of the field of view, for improving the visual comparison between unsheared and sheared conditions.
The two frames correspond to $\gamma_e=0\%$ (green channel) and $\gamma_e=4.2\%$ (magenta channel). As it can be seen, with $y$ the mismatch increases between the two images due to the shearing deformation of the PDMS membrane.
An image velocimetry software (the Matlab algorithm PIVlab \cite{stamhuis2014pivlab}) has been used to quantify the displacement between consecutive frames over the points of a grid. For each couple $i$ of consecutive frames the displacement along $x$ ($\Delta x_i$) and $y$ ($\Delta y_i$) of each point of the grid has therefore been evaluated. The sums over all the displacements along $x$ and $y$, divided by the $y$ coordinate, are reported in figure \ref{fig:cells}(b). As it can be seen, the displacements along $y$ are rather negligible, while the ones along $x$ progressively increases with $y$. 
Local displacements along $y$ visible in Figure \ref{fig:cells}(b) - please note in particular the horizontal band in the center of the panel - are due to local dynamics of cells on top of the membrane deformation displacements. D.d cells are indeed highly motile and can easily move of few micrometers during the shear application timescale ($5$ s). Despite this, these displacements in $\Delta y$ are much smaller than the ones measured in $\Delta x$, thus attesting that cell dynamics is slow compared to the membrane deformation velocity.
For each point of the grid, the displacement can be converted in the corresponding local strain $\gamma$ by dividing over $y$. In the inset of Figure \ref{fig:cells}(b) the shear strain averaged over the FOV ($\gamma_{FOV}$) has been reported after each shear step in order to check its linearity with the applied macro-deformation $\Delta X$. As it can be seen, the linearity holds.

The test with D.d. demonstrates that the location of non-touching cells moves rigidly with the PDMS membrane upon the application of the shear. In order to check if the same applies with a continuous tissue a similar experiment has been repeated with a confluent monolayer of MDCKII cells, imaged in phase contrast with a 10x objective. 
An effective $6.5\%$ shear has been applied to the cell population once full confluence has been reached. A step shear has been applied with a transient from $0\%$ to $6.5\%$ of less than one second. Cellular dynamics and rearrangements are negligible over this timescale. A frame has been recorded before and after the shear application.
 In Figure \ref{fig:cells}(c) the two acquired frames are reported in magenta and green. As for D.d., the bottom of the image has been translated in order to overlap the cells in the region of lower $y$ to better appreciate the effect of increasing $y$.  
 In order to quantify the local deformation due to the applied shear, a space cross-correlation method has been developed in Matlab (see Appendix \ref{app:MDCKIIshear}). For the MDCKII sample, indeed, the PIVlab algorithm used for D.d. is not suitable as the applied step shear is quite large compared to the small steps applied to the D.d. sample.
  The deformation fields along $x$ and $y$ are reported in Fig. \ref{fig:cells}(d). Similarly to the D.d. data, also the cell monolayer is negligibly displaced along $y$ while the displacements along $x$ are proportional to $y$. The corresponding x-averaged displacement along $x$ is reported in the inset of Fig. \ref{fig:cells}(d), where the linearity in $y$ can be better appreciated. The slope of the linear fit performed on the data (red line) corresponds to the effective shear the cell monolayer is subjected to, \textit{i.e.} $6.5\%$. Please note that - due to the motility of confluent MDCKII cells ($\sim 0.1$ \textmu{}m min\textsuperscript{-1}) smaller than the one of D.d. - displacements along $y$ direction in Fig. \ref{fig:cells}(d) are much smaller than the ones reported in Fig. \ref{fig:cells}(b).

\section{Impact of shear upon the dynamics of a confluent cell monolayer}

In order to demonstrate the potential of the presented apparatus, preliminary results are reported on the effect of the shear upon the dynamics of an MDCKII confluent cell monolayer.

At first, a characterization of epithelial cells dynamics upon PDMS at rest has been made. Substrate softness is not expected to impact on cellular behaviour as the elastic modulus of the cured PDMS is order of magnitude higher than the one of the cells. The substrate can be therefore considered undeformable as a normal petri dish. However, a relevant difference between PDMS and plastic resides in the cell-substrate adhesion, as PDMS treated with APTES is more hydrophobic than glass or plastic petri dishes. Consequently, the monolayer dynamics at zero shear on treated PDMS is expected to be stronger than the one on petri dishes, as already documented when cell-substrate interaction is reduced \cite{pricoupenko2024src,ravasio2015regulation}. In order to quantify how the PDMS membrane affects cellular dynamics, a sample has been prepared coating a normal plastic petri dish with a PDMS membrane and plating MDCKII cells on it. For comparison, a control sample has been prepared without PDMS coating.
Live cell imaging has been performed for both the samples in transmission (10x objective, one frame every 5 minutes) and the velocity has been quantified using PIV.
From the obtained velocity field $\mathbf{V}=\left(V_x,V_y\right)$ the average velocity within the field of view has been evaluated as:

\begin{equation}
%\mathbf{v}_{mig}\left(t\right) = \sqrt{< V_x\left(t\right) >_{x,y}^2+< V_y\left(t\right) >_{x,y}^2 }
\mathbf{v}_{mig}\left(t\right) = < V_x\left(t\right) >_{x,y} \mathbf{u_x}+< V_y\left(t\right) >_{x,y}\mathbf{u_y}
\end{equation}

where $\bf{u_x}$ and $\bf{u_y}$ are the unit vectors along $x$ and $y$ respectively, while $<…>_{x,y}$ denotes the average over both $x$ and $y$ within the field of view. Its absolute value $v_{mig}$ describes the collective motion within the monolayer. A second parameter that captures instead the internal agitation of the monolayer is the root mean squared velocity defined as\cite{villa2022non}:

\begin{equation}
v_{RMS}(t) = \sqrt{ <|\mathbf{V}\left(t\right) - \mathbf{v}_{mig}\left(t\right)|^2>_{x,y} }
\end{equation}

\begin{figure}
\includegraphics[width=0.5\textwidth]{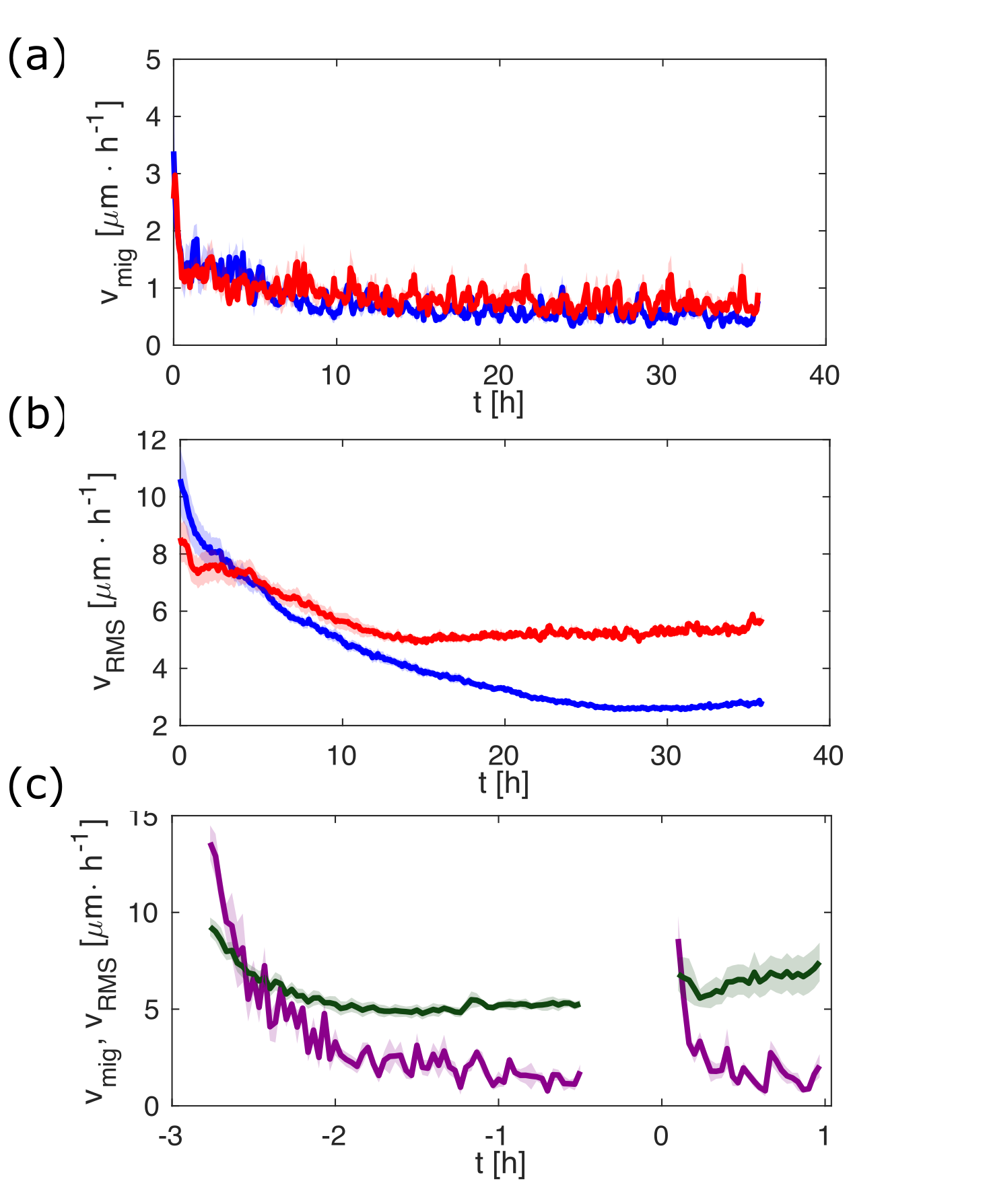}% Here is how to import EPS art
\caption{\label{fig:cellDyn} (a, b) Quantification of the impact of the presence of the PDMS substrate upon the dynamics of a confluent cell monolayer. The dynamic parameters $v_{mig}$ (a) and $v_{RMS}$ (b) are evaluated for the control plastic petri dish (blue) and the PDMS coated petri dish (red). On the horizontal axis, $t=0$ corresponds to the time when the imaging started. Data are averaged over $6$ different fields of view (one replicate). Shadowed regions represent the standard deviation of the mean. (c) $v_{mig}$ (purple) and $v_{RMS}$ (green) evaluated for a MDCKII monolayer before and after a $10 \%$ step shear application. On the horizontal axis, $t=0$ corresponds to the time when shear has been applied. Data are averaged over $5$ different fields of view (one replicate). Shadowed regions represent the standard deviation of the mean.}
\end{figure}

In Figure \ref{fig:cellDyn}(a) and (b) are reported the migration and RMS velocities evaluated for cells in the petri dishes with and without the PDMS layer. All dynamics progressively slow down until they reach a plateau value. The transient dynamics is a consequence of the jamming transition taking place in the monolayer after confluency is reached. The plateau value attests the subsequent steady state of the monolayer dynamics. While no relevant differences between the two samples are documented for the migrating velocity, a significative difference is visible in the plateau value of $ v_{RMS}$. This confirms the expected increase of motility of a confluent monolayer on a low adhesion substrate.

In a following step, cells have been cultured in the PDMS chamber mounted on the shearing device and imaged under the same imaging conditions of the petri dishes. The dynamics has been recorded and characterized with the above mentioned protocol before and after the application of a $10 \%$ step shear. The results for both $v_{mig}$ and $v_{RMS}$ are reported in Figure \ref{fig:cellDyn}(c). The plateau values of both parameters before shear application are comparable with the ones obtained for MDCKII cells on the PDMS coated petri dish. Upon shear application, a sharp increase in $v_{mig}$ can be observed that quickly decays to its pre-shear value. At the same time, a slower but persistent increase in the agitation internal to the monolayer can be observed.
The transient increase of collective motility can be rationalized considering it as a fast reaction of the monolayer to relax the stresses accumulated by the shear. On the other side, the increase of $v_{RMS}$ at longer time scales suggests an increased fluidity of the monolayer. This preliminary observation points towards a two fold impact of shear upon the motility of a confluent cell monolayer, as it appears to differently affect correlated and uncorrelated cellular dynamics.

\section{Conclusions}

The present work introduces a new experimental tool for performing controlled, precise and repeatable shear deformations of 2D soft samples. In particular, the device has been designed for living samples like populations of isolated cells, 2D clusters and confluent cell monolayers. The device enables the live imaging in order to study the effect of shear application over time on the considered sample.
Since the device is compatible with commercial optical microscopes, it can be easily employed in most of the soft matter, biophysics and biology labs.
The device is also compatible with confocal microscopy. This would allow to investigate effect of the applied shear on the 3D cell shape (not explored in the present work). The shear application is indeed not expected to introduce marked out-of-plane deformations, but height fluctuations of cells may still result as a consequence of stress application.

Here, a full characterization of the device is reported. 
Effective strains up to $40 \%$ have been successfully tested without observing the formation of wrinkles. This limit can be easily extended by increasing the pre-stretch applied to the chamber. Tests have also been performed with living cells. Experiments with different cell cultures (isolated D.d. cells and confluent monolayers of MDCKII) revealed the capability of the shearing device to apply a shear to living samples in live cell imaging experiments. 
Preliminary results reveal that the application of in-plane shear to epithelial cell monolayers induces an increase in cellular motility at different timescales. A more systematic investigation of the current phenomena with the described setup will enable a deeper understanding of the impact of shear deformations upon cellular dynamics.
This device promises therefore to be a valuable tool for exploring the effect of shear on the dynamics, morphology, development and phenotypes of cells. The systematic investigation of single cell systems will shed light on possible effects of substrate deformation on cellular dynamics, while tests with epithelial monolayers will allow to verify theoretical predictions on phase transitions on the dynamic state of the system \cite{huang2022shear}.
Differently from stretch, in-plane shear is a transformation that preserves the cell projected area while increasing the cell perimeter. Cell resistance to height fluctuations is therefore not involved. Such deformation will therefore allow a more direct insight into the impact of stresses on actomyosin cortex and cell contractility. These studies will also allow a deeper comprehension of the impact of tangential stresses on the adherent junctions.
Overall, this will enhance our understanding of the interplay between external stresses, active biomechanical response of the cells and the dynamic state of the cell monolayer. The presented shearing device has therefore the potential to make a significant contribution to the unveiling of new mechanobiological and cell migration mechanisms.

\begin{acknowledgments}
We wish to acknowledge Noah Alexander Jarbeau for the help in developing the humidity-temperature-CO\textsubscript{2} sensor and Rohan Yadav for the contact angle measurements.
\end{acknowledgments}

\section*{Author Declarations}
\subsection*{Conflict of Interest}
The authors have no conflicts to disclose.
\subsection*{Author Contributions}
\textbf{S.K. Das:} Formal analysis (supporting); Investigation (supporting); Writing – original draft (supporting); Writing – review
and editing (supporting)
\textbf{L. Mamoyan:} Investigation (supporting); Writing – review and editing (supporting)
\textbf{E. Bodenschatz:} Conceptualization (supporting); Resources (leading); Writing – review and editing (supporting)
\textbf{S. Villa:} Conceptualization (lead); Data curation (lead); Formal analysis (lead); Investigation (lead); Methodology (lead); Visualization (lead); Writing – original draft (lead); Writing – review and editing (lead)

\section*{Data Availability Statement}

Data are available on request from the authors.

\appendix

\section{Supplementary figures on the experimental setup}

\begin{figure}
\includegraphics[width=0.5\textwidth]{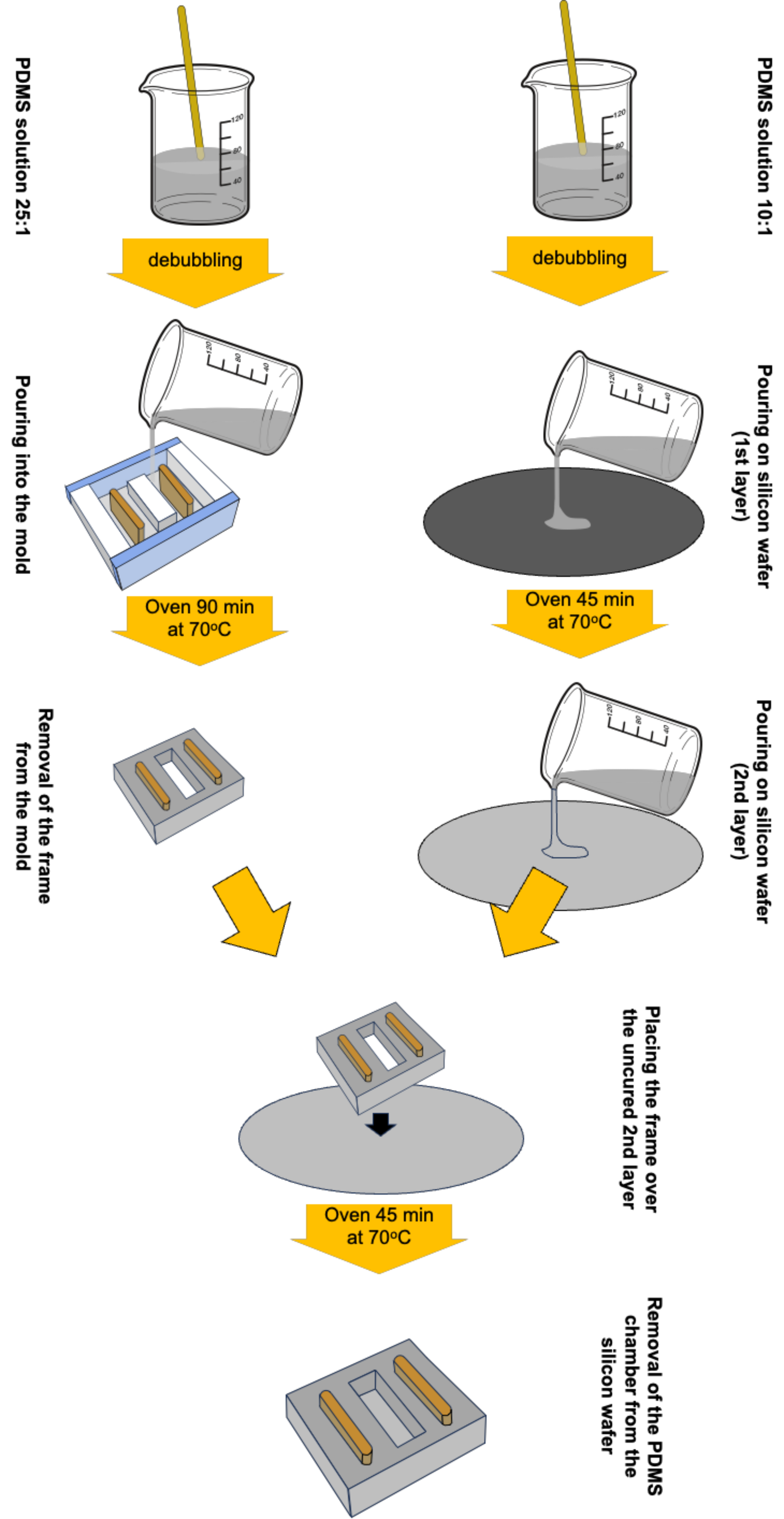}
\caption{\label{fig:PDMSprepSteps} Steps for the production of the PDMS chamber. The PDMS frame (left) and the PDMS membrane (right) are produced in parallel and then joined. Both the frame and the membrane preparations start mixing the base and the curing agent (mixing rations 25:1 and 10:1 respectively). After a de-bubbling step using a vacuum desiccator, both solutions are poured into the mold and upon a silicon wafer respectively. The silicon wafer is spin coated to get a membrane with homogeneous thickness. Both samples are then cured in an oven pre-heated at $70^\circ$C for $90$ (the frame) and $45$ (the membrane) minutes. The PDMS frame is then removed from the mold while a second layer of uncured PDMS is poured upon the first layer on the silicon wafer and spin coated. The frame is then gently placed upon the uncured PDMS layer and the sample is placed in the oven pre-heated at $70^\circ$C for $45$ minutes. Finally, a cutter is used to cut out the shape of the membrane along the edges of the frame and detach the chamber from the silicon wafer.}
\end{figure}

\begin{figure}
\includegraphics[width=0.5\textwidth]{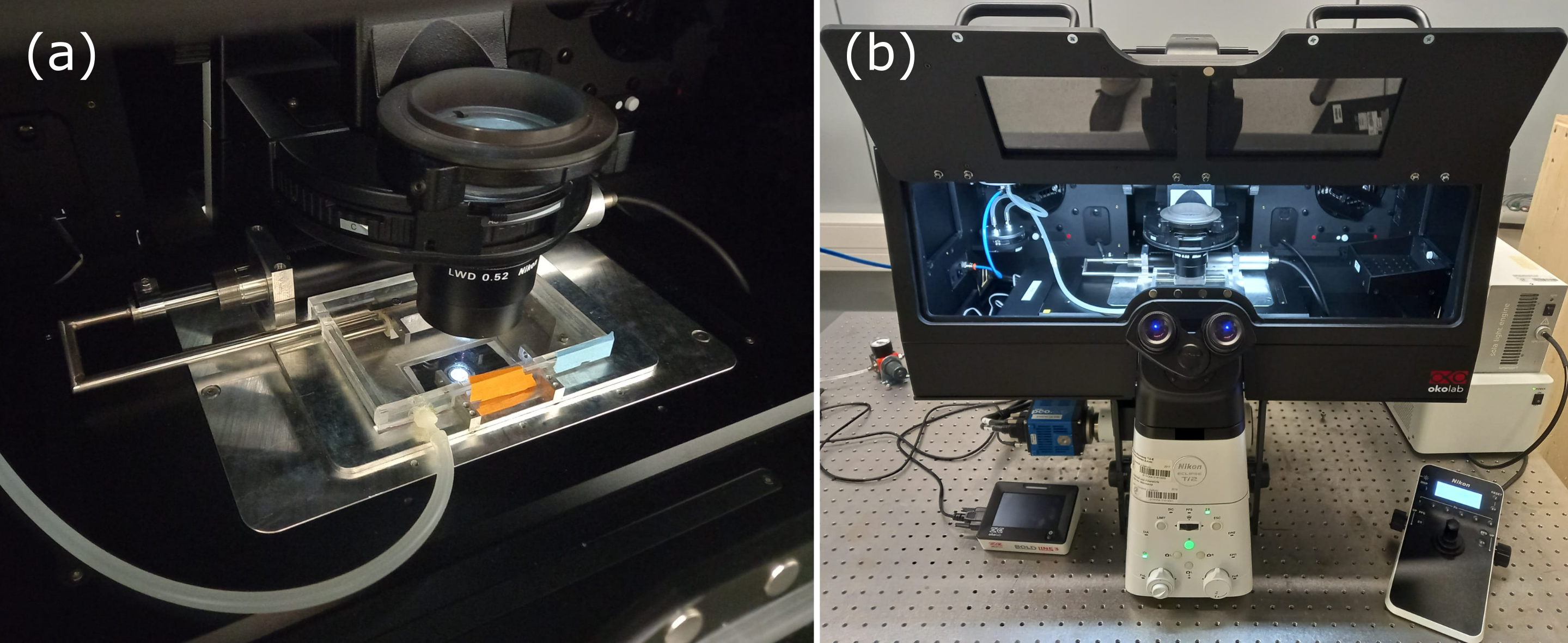}
\caption{\label{fig:nikon} (a, b) Two different views of the shearing device mounted on the stage of a Nikon Eclipse Ti2 inverted microscope mounting an Okolab H203-ENCLOSURE climate chamber.}
\end{figure}

In Figure \ref{fig:PDMSprepSteps} is reported the main steps for the preparation of the PDMS chamber described in the main text.
In Figure \ref{fig:nikon} are reported pictures of the shearing device mounted on a Nikon Eclipse Ti2 inverted microscope mounting an Okolab H203-ENCLOSURE climate chamber.

\section{Strain inhomogeneities}
\label{app:dx_vs_x}

\begin{figure}
\includegraphics[width=0.5\textwidth]{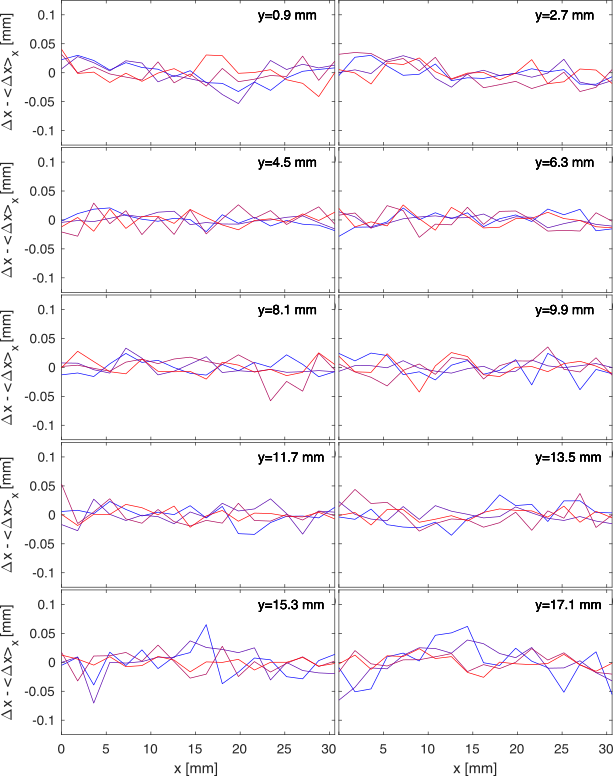}% Here is how to import EPS art
\caption{\label{fig:dx_vs_x} Heterogeneity of deformation $\Delta x\left(x,y\right) - <\Delta x\left(x,y\right)>_x$ along $x$. Each plot refers to a different $y$ coordinate, and each color corresponds to a different applied shear (according to the color sequence of figure \ref{fig:effShear}). A $13\%$ pre-stretch chamber has been used for the characterization.}
\end{figure}

In Figure \ref{fig:dx_vs_x} is reported, for different $y$, the local displacement $\Delta x(x,y)$ corrected by its space average along $x$, in order to appreciate the space dependent deviation of the local displacement. It has been measured by tracking the beads embedded in the PDMS membrane as described in the main text. As it can be seen, the amplitude of the deviations is independent from both the applied strain shear and the value of $y$. It is most likely due to small inhomogeneities in the PDMS membrane. As reported in the main text, for $y>5$ mm such heterogeneity in the displacements translates in a heterogeneity in the effective strain lower than $0.3\%$.

\section{Evaluation of the MDCKII monolayer deformation}
\label{app:MDCKIIshear}

\begin{figure}
\includegraphics[width=0.5\textwidth]{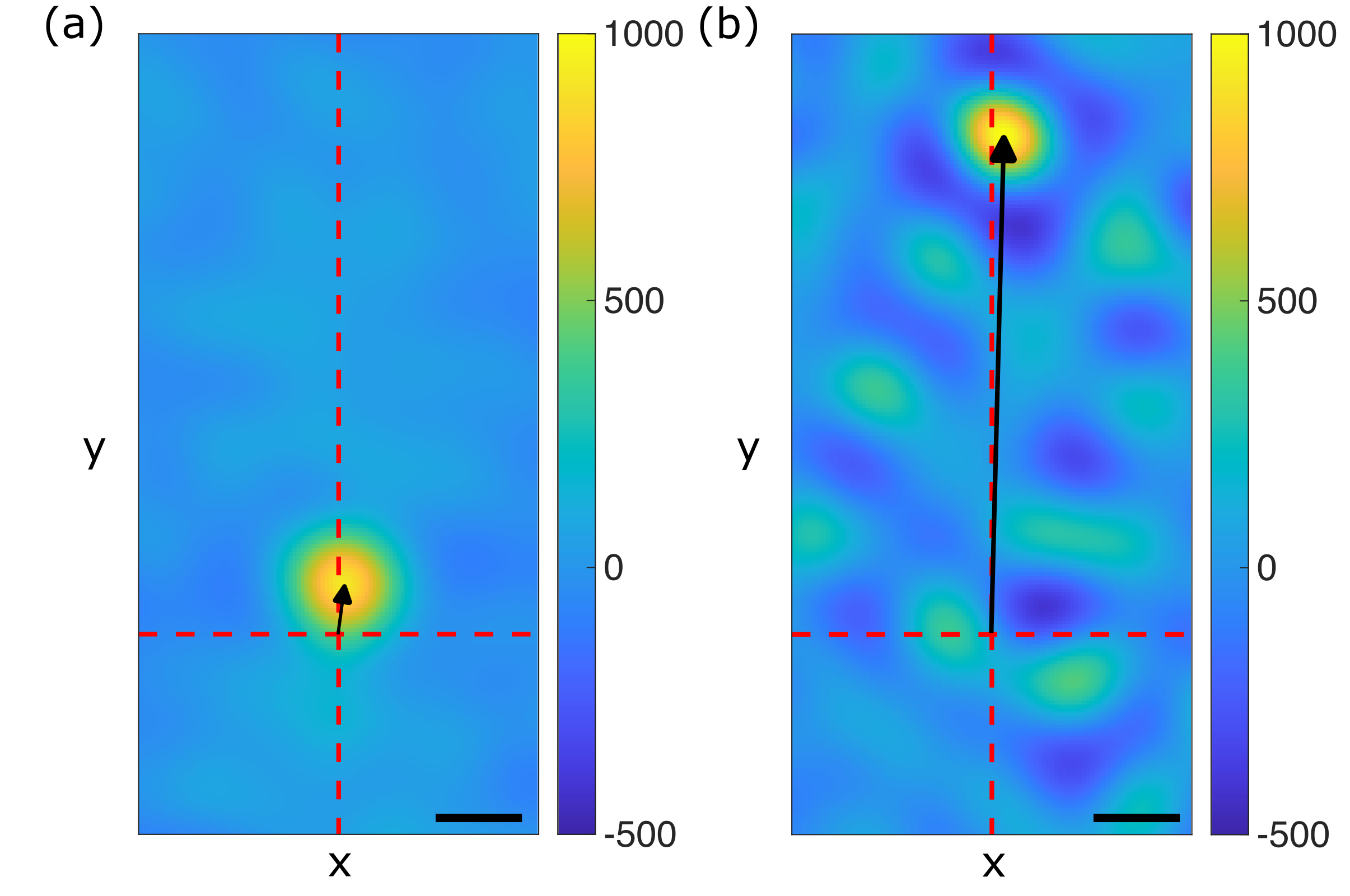}% Here is how to import EPS art
\caption{\label{fig:corr} 2D Space cross correlations computed between unsheared and sheared ROIs corresponding to the bottom right (a) and the top left (b) of the displacement maps reported in Figure \ref{fig:cells}(d). The reported figures are zoom in of the entire correlation matrix, where the red dashed lines represent the zeros axis for $x$ and  $y$. Arrows highlight the relative distance between the origin and the correlation peaks. The arrows components correspond to the average displacement along $x$ and $y$ the monolayer underwent in that specific ROI. Scalebars correspond to $10$ \textmu{}m.}
\end{figure}

In order to quantify the local deformation of the MDCKII monolayer due to the applied shear, a space cross-correlation method has been developed in Matlab. The algorithm operates as it follows. First, both the images of the unsheared ($I_u$) and sheared ($I_s)$ monolayers have been divided in adjacent squares with side $116$ \textmu{}m, equivalent to $250$ pixels. For each couple of ROIs, the following operations have been performed: (i) gaussian filter (width $5$ pixels), (ii) normalization and finally (iii) space cross correlation between the filtered and normalized couples of ROIs. The gaussian filter is applied to remove the smallest subcellular features that add noise to the space correlation. The latter, has been performed using the Matlab function \textit{xcorr2}. The resulting 2D array exhibits a well-defined peak corresponding to the correlation maximum (Fig. \ref{fig:corr}). The $x$ and $y$ coordinates of the maximum are finally identified using the Matlab \textit{max} function: they correspond to the average displacement along the two axes of the cells contained in the considered ROI of $I_u$ in the time interval of shear application. These are the values reported in Figure \ref{fig:cells}(d).

%\nocite{*}
\bibliography{aipsamp}% Produces the bibliography via BibTeX.

\end{document}